\documentclass[sigconf]{aamas}  % do not change this line!

\usepackage{booktabs}
\usepackage{wasysym}

\usepackage{flushend}
\usepackage{array}
\setcopyright{ifaamas}  % do not change this line!
\acmDOI{doi}  % do not change this line!
\acmISBN{}  % do not change this line!
\acmConference[AAMAS'20]{Proc.\@ of the 19th International Conference on Autonomous Agents and Multiagent Systems (AAMAS 2020), B.~An, N.~Yorke-Smith, A.~El~Fallah~Seghrouchni, G.~Sukthankar (eds.)}{May 2020}{Auckland, New Zealand}  % do not change this line!
\acmYear{2020}  % do not change this line!
\copyrightyear{2020}  % do not change this line!
\acmPrice{}  % do not change this line!

\usepackage{color}
\usepackage[textsize=scriptsize]{todonotes}
\definecolor{blue}{RGB}{0, 93, 170}			%Go Big Blue!
\definecolor{darkgreen}{RGB}{0, 102, 0}

\begin{document}

\title{Heuristic Strategies in Uncertain Approval Voting Environments}  % put your title here!
%\titlenote{Produces the permission block, and copyright information}

% AAMAS: as appropriate, uncomment one subtitle line; check the CFP
%\subtitle{Extended Abstract}
%\subtitle{Blue Sky Ideas Track}
%\subtitle{JAAMAS Track}
%\subtitle{Demonstration}
%\subtitle{Doctoral Consortium}

% AAMAS: submissions are anonymous for most tracks
%\author{Paper \#1032}  % put your paper number here!

%% example of author block for camera ready version of accepted papers: don't use for anonymous submissions
%
\author{Jaelle Scheuerman}
%\authornote{Dr.~Trovato insisted his name be first.}
%\orcid{1234-5678-9012}
\affiliation{%
  \institution{Tulane University}
  \streetaddress{6823 St Charles Ave}
  \city{New Orleans} 
  \state{LA} 
  \postcode{70001}
}
\email{jscheuer@tulane.edu}
\author{Jason L. Harman}
%\authornote{The secretary disavows any knowledge of this author's actions.}
\affiliation{%
  \institution{Louisiana State University}
  %\streetaddress{P.O. Box 1212}
  \city{Baton Rouge} 
  \state{LA} 
  %\postcode{43017-6221}
}
\email{jharman@lsu.edu}
\author{Nicholas Mattei}
%\authornote{This author is the
%  one who did all the really hard work.}
\affiliation{%
  \institution{Tulane University}
  %\streetaddress{1 Th{\o}rv{\"a}ld Circle}
  \city{New Orleans} 
  \state{LA}}
\email{nsmattei@tulane.edu}
\author{K. Brent Venable}
\affiliation{
  \institution{Institute for Human \& Machine Cognition}
  \institution{University of West Florida}
  \city{Pensacola}
  \state{FL}
}
\email{bvenabl@ihmc.us}
%\author{Aparna Patel} 
%\affiliation{%
% \institution{Rajiv Gandhi University}
% \streetaddress{Rono-Hills}
% \city{Doimukh} 
% \state{Arunachal Pradesh}
% \country{India}}
%\author{Huifen Chan}
%\affiliation{%
%  \institution{Tsinghua University}
%  \streetaddress{30 Shuangqing Rd}
%  \city{Haidian Qu} 
%  \state{Beijing Shi}
%  \country{China}
%}
%
%\author{Charles Palmer}
%\affiliation{%
%  \institution{Palmer Research Laboratories}
%  \streetaddress{8600 Datapoint Drive}
%  \city{San Antonio}
%  \state{Texas} 
%  \postcode{78229}}
%\email{cpalmer@prl.com}
%
%\author{John Smith}
%\affiliation{\institution{The Th{\o}rv{\"a}ld Group}}
%\email{jsmith@affiliation.org}
%
%\author{Julius P.~Kumquat}
%\affiliation{\institution{The Kumquat Consortium}}
%\email{jpkumquat@consortium.net}
%
%% The example's default list of authors is too long for headers
%\renewcommand{\shortauthors}{B. Trovato et al.}

\begin{abstract}  % put your abstract here!
In many collective decision making situations, agents vote to choose an alternative that best represents the preferences of the group. Agents may manipulate the vote to achieve a better outcome by voting in a way that does not reflect their true preferences. In real world voting scenarios, people often do not have complete information about other voter preferences and it can be computationally complex to identify a strategy that will maximize their expected utility. In such situations, it is often assumed that voters will vote truthfully rather than expending the effort to strategize. However, being truthful is just one possible heuristic that may be used. In this paper, we examine the effectiveness of heuristics in single winner and multi-winner approval voting scenarios with missing votes. In particular, we look at heuristics where a voter ignores information about other voting profiles and makes their decisions based solely on how much they like each candidate. In a behavioral experiment, we show that people vote truthfully in some situations and prioritize high utility candidates in others. We examine when these behaviors maximize expected utility and show how the structure of the voting environment affects both how well each heuristic performs and how humans employ these heuristics.
\end{abstract}

  % put your semicolon-separated keywords here!

\maketitle

%%%%%%%%%%%%%%%%%%%%%%%%%%%%%%%%%%%%%%%%%%%%%%%%%%%%%%%%%%%%%%%%%%%%%%%%%%%%%%%%%%%%%%%%%%%%%%%%%%%%%%%%%
%% start of main body of paper

\section{Introduction}
Computational Social Choice (COMSOC) investigates computational issues surrounding the aggregation of individual preferences and collective decision making \cite{BCELP16a}. Much of this work has focused on the computational complexity of manipulating elections under different voting rules. When it is computationally prohibitive to manipulate an election, it is assumed that voters will vote with their true preferences rather than trying to strategize \cite{faliszewski2010acm}.

Voting truthfully is just one possible heuristic that voters may use when faced with complex voting scenarios. A recent study of voting behavior in multi-winner approval elections showed that the majority of voters did not vote truthfully or optimally \cite{scheuerman2019behec}. Instead, the predominant strategy was to use a \textit{take the X best} heuristic which prioritized the highest utility candidates. Another study showed that in a plurality election where a preferred candidate was dominated, voters would compromise and vote for the leader \cite{TMG15a}. 

The effectiveness of a particular heuristic depends on the environment in which it is being used. Decision science research has examined heuristic decision making in complex and uncertain situations. Sometimes, heuristics are viewed as second best shortcuts, when the environment is too complex to use rational strategies \cite{tversky1974judgment}. However, researchers have also shown that heuristics are adaptive strategies that work in the real world, leveraging natural cognitive abilities that exploit the structure of the environment, often leading to better outcomes with the use of less information. Key to this view of heuristics is that in uncertain environments, decision strategies that ignore some information can sometimes achieve better performance than more complicated optimization strategies in situations that are computationally complex or uncertain \cite{gigerenzer1999fast}, e.g., stock market predictions \cite{demiguel2007optimal}.%\nick{can we give a concrete example -- some math nerd might argue here} \jaelle{How's this?}

This paper examines the effectiveness of heuristics in single winner and multi-winner approval voting elections with uncertainty.  Specifically, voters are presented with scenarios where there are missing votes, contrary to the standard COMSOC assumption of complete information for strategic agents \cite{BCELP16a}. In approval voting, an agent may vote for as many candidates as they wish. Winners are chosen by tallying up the votes and choosing the top-$k$ candidates receiving the most votes. Under the basic approval voting rule, an optimal manipulation can be computed in polynomial time when an agent has complete information about the preferences of all the voters \cite{MPRZ08a}. However, it has been shown that when information about voting preferences is missing, computing the possible winners or manipulating the vote is computationally complex \cite{walsh2007uncertainty}. Although manipulation may be computationally hard, voting truthfully or using other heuristics may still maximize a voters expected utility. 

\subsection{Contribution} We show how specific underlying features of voting environments affect how well different heuristics perform compared to maximizing expected utility. In particular, we look at heuristic strategies including \textit{truthful}, \textit{take the X best}, and \textit{regret minimization} in scenarios where less preferred, neutral and disliked candidates lead in an election. In our context, using a \textit{truthful} strategy means voting for all candidates with positive utility. Voting for the highest utility candidate is an example of the \textit{take the X best} heuristic, with $X=1$. \textit{Regret minimization} refers to voting for all liked and neutral candidates to reduce the chance that a disliked candidate, represented by negative utility, will win the election. In a behavioral experiment of 104 subjects on Mechanical Turk, we show how the number of winners, the current leader(s) and the number of missing votes affect the heuristic strategies that people use in approval voting. We find that in many situations most subjects do not vote completely truthfully or maximize their expected utility. Our work provides key insights on human behavior in voting environments. This can lead to more realistic simulation tools and more accurate predictions of election outcomes when approval voting is used. Our study can also inform the design of automated decision support systems by providing evidence about which heuristics humans may be inclined to use in different contexts and help in designing suggestions which take these behavioral aspects into account. Heuristics adopted by humans can also inspire the design of fast and frugal algorithms for tackling problems of prohibitive size or complexity. 

\section{Preliminaries}
We give a brief overview of the mathematical formalism used to study approval voting and formally define the heuristics that we will consider in this paper.

\subsection{{A}pproval Voting} Following \citet{AGGM+15a} and \citet{Kilg10a} we consider a social choice setting $(N,C)$ where we are given a set $N = \{1, \ldots, n\}$ of voting agents and a disjoint set $C = \{c_1, \ldots, c_m\}$ of candidates.  Each agent $i \in N$ expresses an approval ballot $A_i \subseteq C$ which  gives rise to a set of approval ballots $A = \{A_1, \ldots, A_n\}$, called a profile.  We study the multi-winner approval voting rule that take as input an instance $(A, C, k)$ and return a subset of candidates $W \subseteq C$ where $|W| = k$ called the winning set.  

Approval Voting (AV) finds the set $W \subseteq C$ where $|W| = k$ that maximizes the total weight of approvals (approval score), $AV(W) = \sum_{i\in N} |W \cap A_i|$.  Informally, the winning set under AV is the set of candidates that are approved by the largest number of voters.

In some cases it is necessary to use a tie-breaking rule in addition to a voting rule in order to enforce that the size of $W$ is indeed $k$. Tie-breaking is an important topic in COMSOC and can have significant effects on the complexity of manipulation of various rules even under idealized models \cite{AGM+13a,MNW14a,OEH11a}. Typical in the literature a lexicographic tie-breaking rule is given as a fixed ordering over $C$ and the winners are selected in this order. However, in this paper, as discussed in \citet{AGM+13a}, we break ties by selecting the winner randomly to more closely simulate a real world election. 

In order to align our work with the literature on decision heuristics \cite{gigerenzer1996reasoning} we assume that each agent $i \in N$ also has a real valued utility function $u_i: C \rightarrow \mathcal{R}$.  We also assume that the utility of agent $i$ for a particular set of winning candidates $W \subset C$ is $u_i(W)=\sum_{c \in W} u_i(c)$ (slightly abusing notation). If $W$ is the subset elected by the voting rule we will refer to $u_i(W)$ as agent's $i$'s utility for the \emph{outcome} of the election. 

\subsection{Truthfulness and Sincerity in Approval Ballots}

The literature on approval voting for multi-winner elections goes back at least nearly 40 years to the work of  \citet{brams1980approval}.  For nearly that entire period there has been intense discussion of the strategic aspects of approval balloting \cite{brams1982strategic}.  Researchers over the years have made a variety of assumptions and (re)definitions of what makes a particular vote either \emph{truthful} or \emph{strategic}. Much of this commentary is captured in introductory chapter to the Handbook of Approval voting \cite{LaSa10a}.  As detailed in \citet{LaSa10a}, \citet{Niem84a} quotes the following definition of Sincere Approval Voting from \citet{brams1982strategic}: ``A voter votes sincerely if and only if whenever he votes for some candidate, he votes for all candidates preferred to that candidate'' and writes ``Note that this definition includes nothing about approval as such; it does not require voting only for ‘approved’ alternatives.''  Niemi even writes, ``the existence of multiple sincere strategies almost begs the voter to behave strategically.'' \cite{Niem84a}.

Within the COMSOC community this issue has arisen a number of times: what does it mean to be sincere and/or truthful in a given situation?  This is nicely expressed by \citet{endriss2007vote}, ``In approval voting, a ballot consists of the names of any subset of the set of candidates standing; these are the candidates the voter approves of. The candidate receiving the most approvals wins. A ballot is considered sincere if the voter prefers any of the approved candidates over any of the disapproved candidates. Hence, there will be multiple sincere ballots for any given preference ordering.''

However, this does rest on an assumption about the underlying preference model, as discussed by \citet{MPRZ08a}, when agents are only endowed with binary utilities, a truthful vote is always a strategic vote \cite{MPR08b,MPRZ08a}, i.e., approval voting is incentive compatible. A \emph{strategic} vote is one in which an agent maximizes their total (expected) utility given a particular decision setting. 
%Note that in some cases, e.g., under the standard approval voting rule when agents only have binary utilities, a truthful vote is always a strategic vote \cite{MPR08b,MPRZ08a}.  
However, as \citet{MPRZ08a} continues, ``manipulation in Approval is a subtle issue, since the issue may be ill-defined when the voters are assumed to have linear preferences over the candidates. In this case, there are multiple sincere ballots (where all approved candidates are preferred to all disapproved candidates).''

Given this discussion we make the following distinctions:
\begin{enumerate}
\item In the presence of additive utilities and multiple winners we assume that a \emph{completely truthful} vote is one where the voter approves all candidates for which they have positive utility.

\item A \emph{sincere vote}, which includes the definitions of \citet{endriss2007vote}, \citet{MPRZ08a}, and \citet{brams1982strategic}, is one in which if a voter prefers a particular candidate, then he approves all candidates that are preferred to that particular candidates. Intuitively, this is an assumption of monotonicity over the preferences and captures many of the votes one would cast in the \emph{take the X best} hueristic discussed in the following.
\end{enumerate}

We argue, and will use, the terminology that any vote that is not completely truthful by our definition is considered strategic.  While it is the case that these votes may be sincere, we argue they are not completely truthful as, given the definitions above, it is strategically leaving some information out. In what follows we consider, as does much of the literature, the question of which strategic vote to use, and what internal heuristics one may be using to decide it.

\subsection{Heuristic Decision Models}

Heuristics are strategies, or adaptive shortcuts, that humans use to make decisions. This originated with the idea of bounded rationality \cite{simon1955behavioral}, or the observation that both the human mind and the environment make application of normative decision models impossible. Heuristics have been studied in depth in a number of fields, see \citet{hilbig2012review} for an overview. Though sometimes seen as second best alternatives when maximization is not possible, other research views heuristics as ideally adapted strategies that capitalize on the structure of the environment to provide solutions when optimization is not an option, e.g. NP-hard problems, ill defined problems, or unfamiliar/time sensitive problems. In many situations, heuristics have been shown to outperform solutions that use more complex algorithms (i.e. stock market predictions \cite{demiguel2007optimal}). 
    
%Though this literature has a rich taxonomy of heuristics, their building blocks, and the environments where they perform well, 
Key to this idea of heuristics is that they are fast, composed typically of three steps: search, stop, and apply decision rules. They are also frugal, ignoring some of the casually relevant information. Proponents of fast and frugal heuristics have promoted the idea of ecological rationality which examines the  rationality of a decision strategy in its environment \cite{marewski2011cognitive}.
For example, if a simple heuristic that ignores some causally relevant information provides the best solution in a specific environment (i.e. using regret minimization when a disliked candidate leads the election) it is considered ecologically rational, despite violating axioms of normative theories (such as expected utility theory). 

We present three such heuristics inspired from the literature that we believed a priori could be used in single winner and multi-winner approval voting. These include \textit{truthful}, \textit{take the X best}, and  \textit{regret minimization}. Each of these strategies ignore information (i.e. the total votes so far) and use only the utility of each candidate to decide who to vote for.

\subsubsection{Truthful.}
We define a \emph{truthful} vote as one where an agent approves of all candidates for which they have positive utility. This corresponds to the notion of \emph{completely truthful} above.

\subsubsection{Take the X Best.}
When an agent votes with the \textit{take the X best} heuristic, they vote for a subset of the truthful vote. First, they order the list of candidates by the utility value. Formally, $T = t_1 > ... > t_x$ where $u_1(t_1) > ... > u_X(t_X)$. The agent will then vote for the top-$X$ candidates in the list. $X$ could be calculated using a magnitude cut off or a proportional difference between preferences \cite{brandstatter2006priority}. We do not use specific rule to choose $X$, opting instead to test all sizes of $X$. In this paper, we examine situations where there candidates' utility values are not tied, so no tie-breaking rule is assumed. In the future, it would be interesting to explore if and how voters choose between candidates with equal utility.

\subsubsection{Regret Minimization.}
\textit{Regret minimization} takes into account the voter's anticipated regret if a particular disliked candidate were to win the election. Rather than try maximize their utility, the voter may choose to minimize the chance that the disliked candidate(s) will win by voting for all other candidates, whether they generate positive utility or not \cite{zinkevich2008regret,zeelenberg1999anticipated}.

\section{Related Work}

Approval voting is a set of methods for aggregating group preferences that is particularly popular among economists, computer scientists, psychologists, and beyond \cite{LaSa10a,BrFi07c}.  There are even multiple political action committees (PACs) in the United States, e.g., The Center for Election Science\footnote{\url{https://www.electionscience.org/}}, that are committed to seeing the United States change voting procedures to approval voting.  One reason for this popularity is the idea that participants are allowed to express preference over a set of candidates and not just a single one.  In France, a large study was run parallel to the 2002 election showing that many voters would have preferred approval ballots to traditional plurality ballots \cite{LaVa08a}.

%Academic studies of approval voting (AV) include numerous studies about 
The complexity of manipulation for various types of approval voting (AV) has received considerable attention in the COMSOC literature \cite{BCELP16a}.  Assuming that agents act rationally and have full information about the votes of other agents, when agents have \emph{Boolean utilities}, i.e., when all agents either have utility 1 or 0 for candidates they approve or disapprove of, respectively, AV is strategy-proof.  When agents have general utilities, finding a vote that maximizes the agent's utilities can be computed in polynomial time \cite{MPR08b,MPRZ08a}.  For variants of AV including Proportional Approval Voting, Satisfaction Approval Voting, and the Repeated Approval Voting, the complexity of finding utility maximizing votes ranges in complexity from easy to coNP-complete \cite{AGGM+15a}.

%(see \citet{AGGM+15a} for a full discussion of the computational issues of these variants of approval voting).

Many theoretical works in COMSOC make worst case computational assumptions: manipulators have complete information, all votes are known, etc.  However, there also are several efforts to expand these worst case assumptions and strategic issues to include the presence of uncertain information or when agents are not perfectly rational.  In \citet{ReEn12a}, agents are given access to poll information and agent behaviors are modeled as being $k$-pragmatist, i.e., they only look at the top $k$ candidates when deciding whether or not to make a strategic decision.  In \citet{MLR14a}, agents are modeled as behaving in \emph{locally dominant} ways, i.e., they take into account only a small number of possible outcomes when deciding whether or not to act strategically in a particular voting setting.  A survey of other recent work on issues surrounding strategic voting is given by \citet{Meir18a}.

There is a growing effort to use simulations and real-world data to test various decision making models, e.g., \cite{MaWa17,MaWa13a}. Within the economics and psychology literature there have been several studies of approval voting and the behavior of voters. Perhaps the most interesting and relevant to our work is the studies of \citet{RHT07a} which focus on elections of various professional societies where approval balloting was used and the work of \citet{ZMD15a} where many approval voting settings were obtained from Doodle, an online polling platform.  In \citet{RHT07a} election data is used along with proposed heuristics for individual choice behavior, the conclusion is that many voters use a \emph{plurality heuristic} when voting in AV elections, i.e., they vote as if they are in a plurality election, selecting only their most preferred candidate.  In both of these works only AV with a single winner was investigated and both works relied on real-world elections where it was not possible to tease out the relationships between environment and decision. To our knowledge, the work presented in this paper is the first that examines human voting behavior in multi-winner approval settings. We also build upon this work by introducing a new behavioral experiment that examines how voters select different strategies, depending on the underlying environmental factors (i.e. number of winners and number of missing votes). %We also show how strategies, such as voting truthfully or for a single preferred candidate, can sometimes be effective in maximizing one's expected utility.
 
Three recent papers address strategic voting under the plurality rule where agents are making decisions in uncertain environments.  First, \citet{TySc16a} study the voting behavior of agents under the plurality rule with three options.  They find that the amount of information available to the voters affects the decision on whether or not to vote strategically and that in many cases the strategic decisions do not affect the outcome of the plurality vote. Second, in \citet{TMG15a} an online system is presented where participants vote for cash payments in a number of settings using the plurality rule under uncertainty. Two specific scenarios are studied: one where a user votes after being given access to a large pre-election poll and the second where agents vote simultaneously and can update their votes.  They find that most participants do not engage in strategic voting unless there is a clear way to benefit. In the iterative setting most voters were lazy and if they did vote strategically, they would do a one step look ahead or perform a best response myopically. Finally, in \citet{FLMG19a} a comprehensive study using both past datasets and newly collected ones examines the actual behavior of agents in multiple settings with uncertainty versus behavior that is predicted by a number of behavioral and heuristic models.  The paper proposes a novel model of user voting behavior in these uncertain settings called \emph{attainable utility}, where agents consider how much utility they would gain versus the likelihood of particular agents winning given an uncertain poll.  They conclude that the attainable utility model is able to explain the behavior seen in the experimental studies better than existing models and even perform near the level of state of the art machine learning algorithms in modeling users' actual behavior.
We expand upon this work on plurality to consider heuristics in approval voting environments with uncertainty, showing that it may be ecologically rational for voters to use heuristics over more complex optimization strategies.

\begin{figure}[h]
    \centering 
    \includegraphics[width=0.4\textwidth]{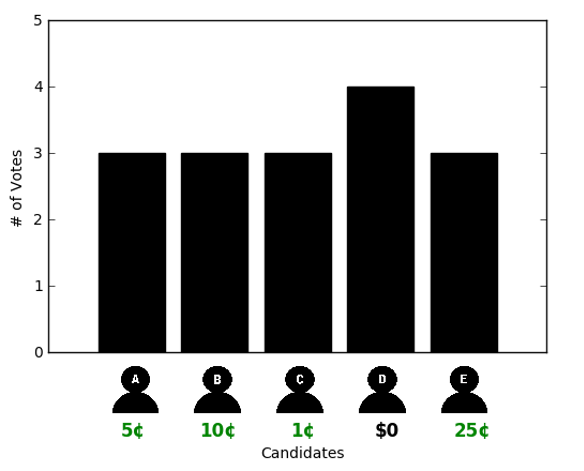}
    \caption{Subjects' view of Scenario 1a details, including the candidates, utility and votes. \textit{Heuristic votes}: Truthful: [A,B,C,E], Take 1 Best: [E], Take 2 Best: [E,B], Take 3 Best: [E,B,A] }
    \label{fig:scenario1a}
\end{figure}

\section{Methods}
%In this section, we describe the design of an experiment used to examine voting behavior in four different scenarios. 

%The goal of this study was to understand if people vote optimally, truthfully, or use some other heuristic. It was also necessary to understand if people's strategies change as features in the underlying environment change.  
To begin, we identify specific scenarios that a voter may encounter in approval voting (Section \ref{section:scenarios}). For each scenario, we explore which strategies people use, if they maximize expected utility, and whether people vote truthfully for all candidates with positive utility or use some other approach. 

Each scenario consists of a set of candidates $C = \{c_1,...,c_5\}$ with $i$'s utility for each candidate in $[-1.0,0.25]$
%\nick{this isn't righ is it -- don't we go up to \$1?}. \jaelle{I think it is right. We only go up to 0.25, where the -\$1 is to represent very strong dislike} 
We manipulate two environmental features, including the number of winners ($k = {1,2, 3}$) and the number of missing votes ($n={0,1, 3}$).
%representing voters whose ballot is not yet known). 
When the final ballots result in a tie, the winner(s) are chosen randomly.

For each scenario, the maximum expected utility can change for different numbers of winners and missing votes.  We calculate the expected utility by generating the power set of all possible votes that $i$ could cast over $C$, i.e. $V = \mathcal{P(}$$C$ $\mathcal{)}$.

The expected utility is then calculated for each of $i$'s votes in $V$ for every combination of $k$ winners, and $n$ remaining voters.

\[E[u(v,k)] = p_1u_i(c_1) + ... + p_mu_i(c_m) \]
\[\forall v \in V, 0 < k < 3, 0 < m < 5 \]

Here, $p_j$ refers to the probability that candidate $c_j$ is in the current profile. For all combinations of numbers of winners and missing votes, we calculate the expected utility for each heuristic, the maximum expected utility, and any votes in $V$ not represented by the heuristics that maximized $i$'s expected utility.

This calculation shows how the computation of an expected utility maximizing strategy is a cognitively demanding option to which  heuristics can be low effort alternatives.

\subsection{Scenarios} \label{section:scenarios}
Below we detail the candidates, utilities, the number of current votes, and the number of missing votes (if applicable) for each candidate for several partial profiles, which we designed to study specific behaviors as the number of winners ($k$) and the number of missing votes ($n$) change.  The scenarios were presented as depicted in Figure \ref{fig:scenario1a} along with a text description of how many voters remained to vote.  When applicable we also show the heuristic strategy that would lead to the maximum expected utility, which we would expect each participant to employ when presented with a particular voting scenario.

\begin{table}[H]
\centering
\begin{tabular}{| m{1em} | m{3em} m{3em} m{3em}|} 
 \hline
  & \multicolumn{3}{c|}{ \# winners ($k$)} \\
 \hline
 $n$ & 1 & 2 & 3 \\
 \hline
 0 & 0.12 \newline \textbf{Take 1} & 0.22 \newline \textbf{Take 1} & 0.31 \newline \textbf{Take 2} \\
 \hline
 1 & 0.11 \newline \textbf{Take 1} & 0.21 \newline \textbf{Take 2} & 0.30 \newline \textbf{Take 2} \\
 \hline
 3 & 0.11 \newline \textbf{Take 1} & 0.20 \newline \textbf{Take 2} & 0.29 \newline \textbf{Take 2} \\
 \hline
\end{tabular}
\caption{Scenario 1a: Maximum expected utility and the voting strategies that achieve it. $n$ represents the number of missing votes.}
\label{tab:scenario1a}
\end{table}
\vspace{-1cm}
\subsubsection{Scenario 1a: Candidate with Trivial Utility}
This scenario (Fig. \ref{fig:scenario1a},
Table \ref{tab:scenario1a})  represents a situation where a non-leading candidate generates a trivial amount of utility if elected. 
%See Figure \ref{fig:scenario1a} and Table \ref{tab:scenario1a} for scenario details.

\begin{table}[H]
\centering 
\begin{tabular}{|c|c|c|c|c|c|}
\hline 
\textbf{Candidate}: & \textbf{A} & \textbf{B} & \textbf{C} & \textbf{D} & \textbf{E} \\
\hline
Utility: & 0.05 & 0.10 & 0.01 & 0.25 & 0 \\
\# Votes: & 3 & 3 & 4 & 3 & 3 \\
\hline
\end{tabular}
\caption{Scenario 1b details, including candidates, utilities and votes. \textit{Heuristic votes}: Truthful: [A,B,C,D], Take 1 Best: [D], Take 2 Best: [D,B], Take 3 Best: [D,B,A]}
\label{tab:scenario1b-utilities}
\end{table}
\vspace{-1cm}
\begin{table}[H]
\centering
\begin{tabular}{| m{1em} | m{3em} m{3em} m{3em}|} 
 \hline
  & \multicolumn{3}{c|}{ \# winners ($k$)} \\
 \hline
 $n$ & 1 & 2 & 3 \\
 \hline
 0 & 0.13 \newline \textbf{Take 1} & 0.26 \newline \textbf{Take 1} & 0.36 \newline \textbf{Take 2} \\
 \hline
 1 & 0.12 \newline \textbf{Take 1} & 0.22 \newline \textbf{Take 2} & 0.31 \newline \textbf{Take 2} \\
 \hline
 3 & 0.11 \newline \textbf{Take 1} & 0.21 \newline \textbf{Take 2} & 0.29 \newline \textbf{Take 2} \\
 \hline
\end{tabular}
\caption{Scenario 1b: Maximum expected utility and voting strategies that achieve it. $n$ represents the number of missing votes.}
\label{tab:scenario1b}
\end{table}
\vspace{-1cm}
\subsubsection{Scenario 1b: Leader with Trivial Utility}
This scenario (Tables \ref{tab:scenario1b-utilities},\ref{tab:scenario1b}) examines a situation where a leading candidate will generate a trivial amount of utility of elected. 
%See Table \ref{tab:scenario1b-utilities} and Table \ref{tab:scenario1b} for scenario details.

\begin{table}[H]
\centering 
\begin{tabular}{|c|c|c|c|c|c|}
\hline 
\textbf{Candidate}: & \textbf{A} & \textbf{B} & \textbf{C} & \textbf{D} & \textbf{E} \\
\hline
Utility: & 0.05 & 0.10 & 0 & 0 & 0.25 \\
\# Votes: & 1 & 1 & 4 & 4 & 1 \\
\hline
\end{tabular}
\caption{Scenario 2a details, including candidates, utilities and votes. \textit{Heuristic votes:} Truthful: [A,B,E], Take 2 Best: [E,B], Take 1 Best: [E]}
\label{tab:scenario2a-utilities}
\end{table}
\vspace{-1cm}
\begin{table}[H]
\centering
\begin{tabular}{| m{1em} | m{3em} m{3em}|} 
 \hline
  & \multicolumn{2}{c|}{ \# winners ($k$)} \\
 \hline
 $n$ & 1 & 2 \\
 \hline
 0 & -- & --  \\
 \hline
 1 & -- & --  \\
 \hline
 3 & 0.01 \newline \textbf{Truth} & 0.04 \newline \textbf{Truth}  \\
 \hline
\end{tabular}
\caption{Scenario 2a: Maximum expected utility and voting strategy that achieve it. $n$ represents the number of missing votes. When $n=0$ and $n=1$, it is impossible to elect a preferred candidate, and all voting strategies lead to an expected utility of 0.}
\label{tab:scenario2a}
\end{table}
\vspace{-1cm}
\subsubsection{Scenario 2a: Dominated for One and Two Winners}
This scenario (Tables \ref{tab:scenario2a-utilities}, \ref{tab:scenario2a}) 
examines a situation where neutral candidates dominate the preferred candidates. When only 1 or 2 candidates can win, there is no possibility of electing a preferred candidate, except when there are 3 missing votes. 
%See Table \ref{tab:scenario2a-utilities} and Table \ref{tab:scenario2a} for scenario details.

% \begin{figure}[ht]
%     \centering 
%     \includegraphics[width=0.5\textwidth]{./images/scenario2a.png}
%     \caption{Scenario 2a details that were shown to participants.}
%     \label{fig:scenario2a}
% \end{figure}

\begin{table}[H]
\centering 
\begin{tabular}{|c|c|c|c|c|c|}
\hline 
\textbf{Candidate}: & \textbf{A} & \textbf{B} & \textbf{C} & \textbf{D} & \textbf{E} \\
\hline
Utility: & 0.10 & 0 & 0 & 0 & 0.25 \\
\# Votes: & 1 & 4 & 4 & 4 & 1 \\
\hline
\end{tabular}
\caption{Scenario 2b details, including candidates, utilities and votes. \textit{Heuristic votes:} Truthful: [A,E], Take 1 Best: [E]}
\label{tab:scenario2b-utilities}
\end{table}
\vspace{-1cm}
\begin{table}[H]
\centering
\begin{tabular}{| m{1em} | m{5.5em} |} 
 \hline
  & \# winners ($k$) \\
 \hline
 $n$ & 3 \\
 \hline
 0 & --  \\
 \hline
 1 & --  \\
 \hline
 3 & 0.05 \newline \textbf{Truth} \\
 \hline
\end{tabular}
\caption{Scenario 2b: Maximum expected utility and voting strategy that achieve it. $n$ represents the number of missing votes. It is impossible to elect a preferred candidate for $n=0$ and $n=1$.}
\label{tab:scenario2b}
\end{table}
\vspace{-1cm}
\subsubsection{Scenario 2b: Dominated for Three Winners}
Like Scenario 2a, this scenario (Tables \ref{tab:scenario2b-utilities},\ref{tab:scenario2b}) examines a situation where neutral candidates dominate the preferred candidates. In this particular scenario there is no possibility of electing a preferred candidate in the 3-winner case. See Table \ref{tab:scenario2b-utilities} and Table \ref{tab:scenario2b} for scenario details.

% \begin{figure}[ht]
%     \centering 
%     \includegraphics[width=0.5\textwidth]{./images/scenario2b.png}
%     \caption{Scenario 2b details that were shown to participants.}
%     \label{fig:scenario2b}
% \end{figure}

\begin{table}[H]
\centering 
\begin{tabular}{|c|c|c|c|c|c|}
\hline 
\textbf{Candidate}: & \textbf{A} & \textbf{B} & \textbf{C} & \textbf{D} & \textbf{E} \\
\hline
Utility: & 0.05 & 0.10 & 0 & -1.00 & 0.25 \\
\# Votes: & 3 & 3 & 4 & 4 & 4 \\
\hline
\end{tabular}
\caption{Scenario 3 details, including candidates, utilities and votes. \textit{Heuristic votes:} Truthful: [A,B,E], Take 1 Best: [E], Take 2 Best: [E,B], Regret Minimization: [A,B,C,E] }
\label{tab:scenario3-utilities}
\end{table}
\vspace{-1cm}
\begin{table}[H]
\centering
\begin{tabular}{| m{1em} | p{3em} p{3em} p{3em}|} 
 \hline
  & \multicolumn{3}{c|}{ \# winners ($k$)} \\
 \hline
 $n$ & 1 & 2 & 3 \\
 \hline
 0 & 0.25 \newline \textbf{Truth, Take 1, Take 2} & 0.25 \newline \textbf{Regret, [C,E]} & -0.03 \newline \textbf{Regret} \\
 \hline
 1 & 0.10 \newline \textbf{Regret.} & 0.06 \newline \textbf{Regret} & -0.10 \newline \textbf{Regret} \\
 \hline
 3 & 0.03 \newline \textbf{Regret} & -0.03 \newline \textbf{Regret} & -0.17 \newline \textbf{Regret} \\
 \hline
\end{tabular}
\caption{Scenario 3: Maximum expected utility and the voting strategies that achieve it. $n$ represents the number of missing votes. [C,E] represents a vote that maximizes expected utility, but does not fall into one of our defined heuristics.}
\label{tab:scenario3}
\end{table}
\vspace{-1cm}
\subsubsection{Scenario 3: Disliked Candidate}
This scenario (Tables \ref{tab:scenario3-utilities},\ref{tab:scenario3}) examines a situation where a candidate will generate negative utility if elected, representing a situation where the voter $i$ dislikes the candidate. 
%See Table \ref{tab:scenario3-utilities} and Table \ref{tab:scenario3} for scenario details.

% \begin{figure}[ht]
%     \centering 
%     \includegraphics[width=0.5\textwidth]{./images/scenario3.png}
%     \caption{Scenario 3 details that were shown to participants.}
%     \label{fig:scenario3}
% \end{figure}

\begin{table}[h!]
\centering 
\begin{tabular}{|c|c|c|c|c|c|}
\hline 
\textbf{Candidate}: & \textbf{A} & \textbf{B} & \textbf{C} & \textbf{D} & \textbf{E} \\
\hline
Utility: & 0.10 & 0 & 0.15 & 0.20 & 0 \\
\# Votes: & 3 & 4 & 3 & 3 & 3 \\
\hline
\end{tabular}
\caption{Scenario 4 details, including candidates, utilities and votes. \textit{Heuristic votes:} Truthful: [A,C,D], Take 1 Best: [D], Take 2 Best: [C,D]}
\label{tab:scenario4-utilities}
\end{table}
\vspace{-1cm}
\begin{table}[h!]
\centering
\begin{tabular}{| m{1em} | m{3em} m{3em} m{3em}|} 
 \hline
  & \multicolumn{3}{c|}{ \# winners ($k$)} \\
 \hline
 $n$ & 1 & 2 & 3 \\
 \hline
 0 & 0.11 \newline \textbf{Truth} & 0.23 \newline \textbf{Truth} & 0.32 \newline \textbf{Take 2} \\
 \hline
 1 & 0.11 \newline \textbf{Truth} & 0.22 \newline \textbf{Take 2} & 0.31 \newline \textbf{Truth} \\
 \hline
 3 & 0.11 \newline \textbf{Take 2} & 0.21 \newline \textbf{Truth} & 0.31 \newline \textbf{Truth} \\
 \hline
\end{tabular}
\caption{Scenario 4: Maximum expected utility and voting strategies that achieve it. $n$ represents the number of missing votes.}
\label{tab:scenario4}
\end{table}
\vspace{-1cm}
\subsubsection{Scenario 4: Neutral Leader}
This scenario (Tables \ref{tab:scenario4-utilities},\ref{tab:scenario4}) examines a situation where a neutral candidate is leading the election. %See Table \ref{tab:scenario4-utilities} and Table \ref{tab:scenario4} for scenario details.

% \begin{figure}[ht]
%     \centering 
%     \includegraphics[width=0.5\textwidth]{./images/scenario4.png}
%     \caption{Scenario 4 details that were shown to participants.}
%     \label{fig:scenario4}
% \end{figure}

\subsection{Experiment Implementation}
\textbf{Participants.} 104 participants were recruited through Mechanical Turk to participate in the voting heuristics study. Participants were paid \$1.00 to complete the survey. They also received a bonus of no more than \$8.00 that was determined by the outcome of the hypothetical elections.

\textbf{Procedure.} In the study, participants were asked to vote in a series of unrelated hypothetical elections. All participants voted in the single winner scenarios (n=104). Participants were then randomly assigned to be part of a 2-winner (n=50) or 3-winner(n=54) election for the remainder of the study. 
%All participants repeated each scenario in situations of increasing uncertainty, first voting in scenarios when there were 0 missing votes, then 1 missing vote and finally, 3 missing votes. 

Participants were asked to give informed consent and then proceeded to study. They read instructions which explained approval voting and the tie-breaking mechanism with examples. After reading the instructions, they proceeded through single-winner scenarios, first encountering scenarios with 0 missing votes, then 1 and finally, 3 missing votes. From there, the survey presented each participant with a series of multi-winner scenarios for their assigned group (2 or 3-winner), in order of increasing uncertainty.

Each election displayed an image showing the candidates, the number of votes cast for each candidate so far, and how much money the participant would earn for each candidate if they were elected. Figure \ref{fig:scenario1a} is an example of what the participants saw. 

When voting, subjects could vote for 0 or more (up to five) of the five candidates. After voting, they would see the election results, including the winners, the amount earned, and the ballots cast by any missing voters (when applicable). 

The experiment was designed so that participants could choose to vote truthfully (for all candidates with positive utility) or manipulate their vote to achieve a higher utility. We expected that most people would try to vote strategically, but since the situations involve varying degrees of uncertainty and were cognitively complex, participants would not perform all of the necessary computations to identify the strategy that maximizes their utility. Instead, we expected that people would use heuristics, such as being \emph{truthful} or using \emph{take the X best}, to prioritize the highest priority candidates.

\section{Results \& Discussion}
The results of the behavioral experiment described above showed unique patterns of behavior in each scenario, particularly across the different conditions. The next few sections describe the results for each scenario.

\begin{figure}[ht]
    \centering 
    \includegraphics[width=0.4\textwidth]{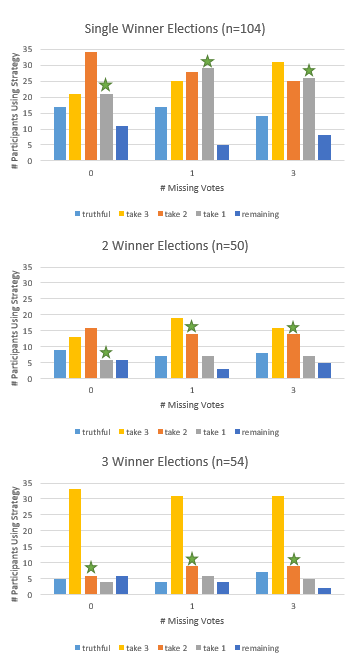}
    \caption{Scenario 1a Results. Maximizing strategies are marked with a star.}
    \label{fig:scenario1a-results}
\end{figure}

\subsection{Scenarios 1a, 1b: Trivial Utilities}  \label{section:scenario1-results}
In these scenarios, we wanted to see whether or not people would vote for a candidate with a trivial utility (represented as a candidate that would earn $1\cent$ if elected). Scenario 1a examined the case when the trivial candidate was not leading the election and Scenario 1b examined when it was. We found that in both scenarios, people generally did not vote truthfully for all candidates with positive utility, including the trivial candidate. In Scenario 1a, only 15.4\% voted truthfully in the 1-winner election, 16.0\% in the 2-winner election, and 9.9\% in the 3-winner election. Scenario 1b was similar with only 14.7\% voting truthfully in the 1-winner election, 11.3\% in the 2-winner election and 8.0\% in the 3-winner election. 

\begin{figure}[ht]
    \centering 
    \includegraphics[width=0.4\textwidth]{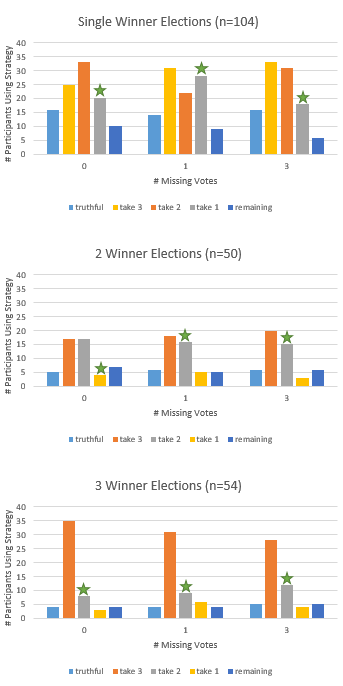}
    \caption{Scenario 1b Results. Maximizing strategies are marked with a star.}
    \label{fig:scenario1b-results}
\end{figure}

In both of these scenarios, it was optimal to use a \emph{take the X best} approach, with $X$ increasing as the number of winners and missing votes increased (see Tables \ref{tab:scenario1a} and \ref{tab:scenario1b}). Although the majority of people used a \emph{take the X best} strategy (Scenario 1a: 77.8\%, Scenario 1b: 78.8\%), they rarely prioritized the $X$ candidates that would maximize the expected utility. In Scenario 1a, only 21.5\% of participants chose a strategy that would lead to an optimal outcome. In Scenario 1b, only 18.4\% chose a maximizing strategy. 

Using $\chi^2$ analysis, we found no significant difference in how people voted as the number of missing votes increased, even in the 2-winner elections where increased uncertainty led to a different maximizing strategy (take the 1 best for 0 missing votes vs. take the 2 best for 3 missing votes). However, significant differences (P < 0.005) were found when comparing the voting strategies used by those electing one or two winners compared to those electing three winners. In general, when voting in the 1-winner and 2-winner elections, participants voted for 2 or 3 candidates (1-winner: 57.9\%, 2-winner: 70.7\%) more often than other strategies. When participants voted in the 3-winner election, they usually voted for 3 candidates (61.7\%) (see Figures \ref{fig:scenario1a-results} and \ref{fig:scenario1b-results}).

%\begin{figure}[ht]
%    \centering 
%    \includegraphics[width=0.5\textwidth]{./images/scenario2a-results.png}
%    \caption{Scenario 2a Results. Maximizing strategies are marked with a star.}
%    \label{fig:scenario2a-results}
%\end{figure}

\subsection{Scenarios 2a, 2b: Dominated Preferences} \label{section:scenario2-results}
In these scenarios, we wanted to see if people would vote truthfully when neutral candidates dominated their preferred candidates. Scenario 2a examined this in the context of 1 and 2-winner elections, where Scenario 3a looked at 3-winner contexts. In both Scenario 2a and 2b it was possible to elect a preferred candidate when there were 3 missing votes, where the maximizing strategy was to vote truthfully (see Tables \ref{tab:scenario2a} and \ref{tab:scenario2b}). Voting truthfully was also the participants' dominant strategy no matter the numbers of winners or missing votes (Scenario 2a: 44.2\%, Scenario 2b: 62.3\%). The second most common strategy was to abstain (Scenario 2a: 16.5\%, Scenario 2b: 20.4\%). 
%(see Figures \ref{fig:scenario2a-results} and \ref{fig:scenario2b-results}).

%\begin{figure}[ht]
%    \centering 
%    \includegraphics[width=0.5\textwidth]{./images/scenario2b-results.png}
%    \caption{Scenario 2b Results. Maximizing strategies are marked with a star.}
%    \label{fig:scenario2b-results}
%\end{figure}

Using $\chi^2$ analysis, we found a significant difference (P < 0.0005) when comparing the voting strategies that people used in Scenario 2a when voting for one winner versus two winners. There was also a significant difference in voting strategies when when there was 0 or 1 missing vote, compared to 3. When there were 0 missing votes, participants chose to abstain 22.7\%, but when there were 3 missing votes, only 2.6\% of participants abstained. This seems to indicate that voters in Scenario 2a recognized that they had a small chance to elect a preferred candidate in the 3-winner condition and voted accordingly.  In Scenario 2b, the number of abstentions decreased as the level of uncertainty increased (0 missing votes: 25.9\% abstain, 3 missing votes: 14.8\% abstain), but it was not enough to result in a significant difference in each groups' voting strategies.

%\begin{figure}[ht]
%    \centering 
%    \includegraphics[width=0.5\textwidth]{./images/scenario3-results.png}
%    \caption{Scenario 3 Results. Maximizing strategies are marked with a star.}
%    \label{fig:scenario3-results}
%\end{figure}

\subsection{Scenario 3: Disliked Candidate} \label{section:scenario3-results}
In this scenario, we explored how people would vote in the presence of a disliked candidate that would generate negative utility if elected. Here, \textit{regret minimization} was a maximizing strategy for most combinations of numbers of winners and missing votes. However, in the single-winner election with 0 missing votes, it was possible to achieve the optimal strategy both by being truthful or using \emph{take the X best}. When one vote was missing in the single winner election, it was best to be truthful. Voting [C,E], a strategy that did not align with any of the heuristics defined in this paper, was also a maximizing strategy in the 2-winner scenario with 0 missing votes. 

This scenario was interesting as people's voting strategy changed significantly (P < 0.005) when comparing the strategies used by those voting in 1-winner elections with 0 or 1 missing votes, to those voting in elections missing 3 votes. In all three single winner groups, more people responded with a truthful (0 missing votes: 29.8\%, 1 missing votes: 36.5\%, 3 missing votes: 30.8\%) or take the 1 best strategy (0 missing votes: 46.1\%, 1 missing votes: 35.6\%, 3 missing votes: 23.1\%), than any other strategy. However, the number of voters using regret minimization (0 missing votes: 5.8\%, 1 missing votes: 4.8\% and 3 missing votes: 17.3\%) increased so that it was the 3rd most popular strategy when 3 voters were missing. 
%(see Figure \ref{fig:scenario3-results}).

The responses to the 2-winner election were more variable, with maximizing strategies being more popular than other strategies (0 missing votes: 38.0\%, 1 missing vote: 48.0\%, 3 missing votes: 50.0\%), but still not used by a majority of the candidates. In the 3-winner election, being truthful was the most popular response, whereas the optimal strategy (regret minimization) was used only 20.4\% of the time.

In general, it was common for participants in this scenario to vote for as many candidates as there were winners in the election. When voting in the 1-winner election, participants voted for one candidate 37.9\% of the time. In the 2-winner election, voting for two candidates was also the most common (40.0\%), and participants in the 3-winner election mostly voted for three candidates (53.7\%).

\begin{figure*}[ht]
    \centering 
    \includegraphics[width=0.9\textwidth]{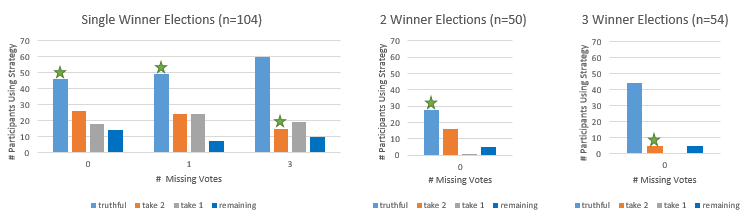}
    \caption{Scenario 4 Results. Maximizing strategies are marked with a star.}
    \label{fig:scenario4-results}
\end{figure*}

\subsection{Scenario 4: Neutral Leading Candidate} \label{section:scenario4-results}

In this scenario, tested if people would vote truthfully when a neutral candidate is leading, even in situations when \emph{take the 2 best} was the maximizing strategy, e.g., when there is 1 winner with 3 missing votes, or 3 winners with 0 missing votes. For this scenario, participants voted in single winner elections with 0, 1 or 3 missing votes, and in 1- 2- or 3-winner elections with 0 missing votes.

We found that when there were 0 missing votes, people's strategies changed significantly (P < 0.005) depending on the number of winners in the election. Overall, being truthful dominated the other strategies (1-winner:49.7\%, 2-winner: 56\%, 3-winner: 81.5\%), especially in the 3-winner election, even though it would have been better to use take the 2 best in this instance. In fact, \emph{take the 2 best} represented only 9.2\% of votes in the 3-winner election. There was no significant difference in people's strategies as the number of missing votes increased. Being truthful was the dominant strategy, even in the 3-winner election, where using take the 2 best had a higher expected utility (0 missing votes: 44.2\%, 1 missing votes: 47.1\%, 3 missing votes: 57.7\%) (see Figure \ref{fig:scenario4-results}).

\subsection{General Discussion}

Behavioral results showed some distinct patterns of voting across all scenarios. The majority of participants \emph{did not vote using a strategy that maximized expected utility}, especially in the 1-winner (25.6\%) and 2-winner (38.4\%) conditions. In the 3-winner condition, 49.6\% voted using a maximizing strategy. We also found that \emph{as the number of possible winners increased, participants were more likely to vote truthfully}, i.e., for all candidates with positive utility (1-winner: 33.6\%, 2-winner: 33.6\%, 3-winner: 46.1\%). We also found that when participants were not entirely truthful, they still tended to use a \emph{take the X best} heuristic, and this captured a significant portion of their responses (1-winner: 50.6\%, 2-winner: 43.8\%, 3-winner: 34.4\%).

We found that people generally used different heuristics in different scenarios and as the numbers of winners changed. For example, in Scenarios 1 (trivial utilities, see \ref{section:scenario1-results}) and 3 (disliked candidate, see \ref{section:scenario3-results}), a significant portion of voters did not vote completely truthfully, and chose to use another strategy such as \emph{take the X best} or \emph{regret minimization}. Voters in these scenarios also tended to vote for a  number of candidates equal to the number of winners they were electing, indicating that they were choosing a heuristic that aligned with the number of winners. However, in Scenarios 2 (dominated preferences, see \ref{section:scenario2-results}) and 4 (neutral leader, see \ref{section:scenario4-results}), being truthful was the dominant strategy by a wide margin, and there was no relation between the number of candidates voted for and the number of winners. 

We found that people were not very sensitive to changes in uncertainty. In Scenarios 1 and 4, \emph{participants' behavior did not significantly change as the number of missing votes increased, even when this resulted in using a non-optimal strategy}. In Scenario 2, voters in the 2-winner elections were sensitive to the fact that they had some chance of electing a candidate when there were 3 missing votes, leading to fewer abstentions in that condition. In Scenario 3, some voters were able to identify that the underlying optimal strategy changed, increasing the number of voters using \emph{regret minimization} from 5.8\% when there were 0 missing votes to 17.3\% when there were 3 missing votes.

\section{Conclusions and Future Work}
%In this paper, we showed that that being \textit{truthful} maximized expected utility in many approval voting scenarios, but also found that other heuristics such as \textit{take the X best} and \textit{regret minimization} maximized expected utility when underlying environmental features, such as the number of winners, the current leading candidate and the number of missing votes were in specific configurations

In this paper we study heuristics in the context of multi-winner elections using approval voting. Our behavioral results show that people do not vote completely truthfully in some approval voting scenarios, such as when there was a candidate with trivial utility or negative utility. In other situations, such as when neutral candidate were leading, people tend to be completely truthful. When people do not vote completely truthfully, they tend to vote sincerely, using a \emph{take the X best} heuristic and they generally are not very effective at choosing the heuristic that maximized their utility. It would be interesting to explore other scenarios in approval voting to see how well these behaviors generalize to other situations.

While the results presented in this paper provide insights into the use and effectiveness of certain heuristics in approval voting, there are many other voting rules and heuristics. It would be interesting to continue exploring heuristics under other voting rules, including those that are known to be computationally complex to manipulate with complete information, such as the single transferable vote (STV). In decision science, taxonomies have been created to show which heuristics may be more or less useful in which environment \cite{gigerenzer2011heuristic}. We believe that a similar approach could prove beneficial to our understanding of voting heuristics which is important for factoring them into a more realistic analysis of the voting rules.

%\begin{acks}

%\end{acks}

%%%%%%%%%%%%%%%%%%%%%%%%%%%%%%%%%%%%%%%%%%%%%%%%%%%%%%%%%%%%%%%%%%%%%%%%%%%%%%%%%%%%%%%%%%%%%%%%%%%%%%%%%
%% bibliography: see CFP for number of permitted pages

\bibliographystyle{ACM-Reference-Format}  % do not change this line!
\bibliography{abb,voting,heuristics}  % put name of your .bib file here

\end{document}